# AUTOMATIC SURVEY-INVARIANT CLASSIFICATION OF VARIABLE STARS

Patricio Benavente[1], Pavlos Protopapas[2], and Karim Pichara[1,3,2]
[1]Computer Science Department, School of Engineering, Pontificia Universidad Católica de Chile, Santiago, Chile
[2]Institute of Applied Computational Science, Harvard University, Cambridge, MA, USA and
[3]Millennium Institute of Astrophysics, Santiago, Chile
*Draft version November 22, 2017*

## ABSTRACT

Machine learning techniques have been successfully used to classify variable stars on widely-studied astronomical surveys. These datasets have been available to astronomers long enough, thus allowing them to perform deep analysis over several variable sources and generating useful catalogs with identified variable stars. The products of these studies are labeled data that enable supervised learning models to be trained successfully. However, when these models are blindly applied to data from new sky surveys their performance drops significantly. Furthermore, unlabeled data becomes available at a much higher rate than its labeled counterpart, since labeling is a manual and time-consuming effort. Domain adaptation techniques aim to learn from a domain where labeled data is available, the *source domain*, and through some adaptation perform well on a different domain, the *target domain*. We propose a full probabilistic model that represents the joint distribution of features from two surveys as well as a probabilistic transformation of the features between one survey to the other. This allows us to transfer labeled data to a study where it is not available and to effectively run a variable star classification model in a new survey. Our model represents the features of each domain as a Gaussian mixture and models the transformation as a translation, rotation and scaling of each separate component. We perform tests using three different variability catalogs: EROS, MACHO, and HiTS, presenting differences among them, such as the amount of observations per star, cadence, observational time and optical bands observed, among others.

*Subject headings:* methods: data analysis – methods: statistical – stars: statistics – stars: variables: general

## 1. INTRODUCTION

Machine learning methods have been applied with success to astronomy problems such as classification of galaxy morphology (Freed & Lee 2013), spectral classification (Christlieb et al. 2002; Bromley et al. 1998), photometric classification (Cavuoti et al. 2014; Brescia et al. 2013), solar activity prediction (Colak & Qahwaji 2009), photometric redshift regression (Benitez 2000; Collister & Lahav 2004), relationship discovery (Graham et al. 2013), anomaly detection (Nun et al. 2014) and variable star classification (Pichara & Protopapas 2013; Pichara et al. 2016; Blomme et al. 2011; Richards et al. 2011), which is the main focus of the technique presented in this paper.

A primary factor that enabled this accomplishment is the availability of data, which is increasing in an accelerating manner. Ever-larger telescopes are being built, and optical sensor sensitivity increases every year. The advent of digital surveys, automated telescopes and on-line catalogs brought astronomy to the big data era. The Sloan Digital Sky Survey (SDSS)[1], designed in 1990, surveyed on the visible spectrum one-third of the sky obtaining positions and brightness of a billion stars, galaxies, and quasars, as well as the spectra of a million objects. Still active today, it generates around 200 GB of data every night, accumulating more than 50 TB of data to date (Feigelson & Babu 2012). The Large Synoptic Survey Telescope (LSST), currently under construction in Chile, is expected to generate an average of 15 Terabytes of data per night upon entering operations in 2022 (Jurić et al. 2015).

As a result, not only more data is available, but each dataset comes from a different survey with distinct characteristics. Indeed, filters and atmospheric conditions vary, and observations differ due to sensor sensitivity and the depth observed in each survey, among other factors. For example, a deep survey may be more biased towards AGNS than a shallower survey. This means that models trained in one survey cannot be readily used in data generated from other surveys and must be retrained from scratch. Moreover, labeled data is unavailable for new surveys, since labeling must be done manually by trained astronomers in a time consuming effort (Sterken & Jaschek 2005). Since applying a previously trained model to a new survey results in considerable losses in performance, this renders supervised learning on new datasets unfeasible.

The latter problem arises from the assumption, often taken in traditional learning techniques, that the distribution of the data used to train the model is the same as the distribution of the data to which the model is applied to. However, this assumption does not generally hold in practical applications.

Therefore, it is desirable to transfer the information learned by a classifier in a domain where labels abound — the source domain — to a domain where few or none labels are available — the target domain. This problem is known as *domain adaptation* (Jiang 2008) and is part of the more general problem of *transfer learning* (Pan & Yang 2010; Raina et al. 2007).

[1] http://www.sdss.org/

In this work, we address the problem of domain adaptation in the context of variable star classification. As such, the source domain is a well-known astronomical survey — in which a relatively high amount of labels exist and a trained classifier performs accurately — and the target domain is a newer or relatively less studied survey where no or very few labels exist.

To solve this, one may use the target dataset instances to induce a change in the source domain classifier that allows it to perform better in the target domain. This approach relies on the creation of an adaptation objective that effectively reduces the classification error on the target domain using no label information. On the other hand, one might find a transformation between the feature spaces of the source and target domains, which allows passing the instances in the target dataset to a representation suitable for training a new classifier. Moreover, this transformation can also be applied in the opposite direction; i.e. to transfer the labeled instances from the source domain to the feature space of the target domain and then train a classifier on this data. We favor the latter approach, as it is model independent. A classifier modification would generally depend on a model's particularities — such as the way a discriminative classifier models class boundaries — while a feature space transformation has the advantage of being model agnostic, decoupling the adaptation and classification problems and thus allowing for the use of the best suited model in a given situation and applying new models as they become available.

We propose a new probabilistic model, based on the Gaussian mixture model (GMM). We use two GMMs to represent the feature distributions of the source and target domains. We then infer linear transformations of the GMM components. We assume that the statistical descriptor shift between the surveys can be corrected by translating, rotating and scaling the GMM components. Our method is unsupervised, as we only require unlabeled data in both domains. We estimate and apply a transformation to the labeled instances in the source domain for each of the mixture components, weighted by how much importance each component has on each data point. In this way, we build a training set suitable for classifying in the target domain. In doing this, we assume that the transformation that corrects the shift in the unlabeled dataset will also correct it in the training set.

Our approach offers some advantages compared to previous research:

1. It finds a transformation from the feature space of one domain to the other, meaning that any data from one domain can be used as if belonging to the other. Other methods perform adaptation at the model level and leave data intact.

2. Since it makes no assumptions about the classifier, our approach is classifier agnostic. Transformed training sets can be used with any model of choice, effectively decoupling domain adaptation from model learning.

## 2. PROBLEM DESCRIPTION AND NOTATION

We follow a notation similar to Jiang (2008). Let $\mathcal{X}$ be the feature space and $\mathcal{Y}$ the label space in our problem. Let $X \in \mathcal{X}$ and $Y \in \mathcal{Y}$ be random variables representing the observed features and the observed class labels, respectively. We denote their true underlying joint distribution as $P(X,Y)$. We distinguish two domains: a source domain where a large amount of labeled data is available and a target domain where labeled data is unavailable or scarce. We denote the true joint distributions of $X$ and $Y$ in the source and target domains as $P_s(X,Y)$ and $P_t(X,Y)$, respectively. Consequently, we denote the true marginal distributions of $X$ and $Y$ for each domain as $P_s(X)$, $P_s(Y)$, $P_t(X)$ and $P_t(Y)$, and the true conditional distributions as $P_s(X|Y)$, $P_s(Y|X)$, $P_t(X|Y)$ and $P_t(Y|X)$, as one would expect.

Let $D_s^l = \{(x_i^{sl}, y_i^{sl})\}_{i=1}^{N_s^l} \subseteq \mathcal{X} \times \mathcal{Y}$ be the labeled data available in the source domain and $D_s^u = \{x_i^{su}\}_{i=1}^{N_s^u} \subseteq \mathcal{X}$ the available unlabeled data in the source domain. Similarly, let $D_t^u = \{x_i^{tu}\}_{i=1}^{N_t^u} \subseteq \mathcal{X}$ be the unlabeled data available in the target domain and $D_t^l = \{(x_i^{tl}, y_i^{tl})\}_{i=1}^{N_t^l} \subseteq \mathcal{X} \times \mathcal{Y}$ the labeled data in the target domain. We call a value $x \in \mathcal{X}$ an unlabeled instance and a tuple $(x,y) \in \mathcal{X} \times \mathcal{Y}$ a labeled instance.

Three types of domain adaptation problems are distinguished based on the kind of data available (Pan & Yang 2010): (a) *Supervised domain adaptation* exploits labeled data both in the source and in the target domain, (b) *unsupervised domain adaptation* uses only unlabeled data, and (c) *semi-supervised domain adaptation* employs only a small amount of labeled data from the target domain.

We focus on the unsupervised domain adaptation problem, therefore we will generally ignore the labeled data in the target domain, $D_t^l$, and use it for testing purposes only.

In our problem, the true joint distributions differ between the two domains: $P_s(X,Y) \neq P_t(X,Y)$. Additionally, several different scenarios can be considered under the domain adaptation problem depending on the assumptions made about the cause of the joint distribution difference between the domains.

### 2.1. Covariate Shift

If we assume that the causal relationships between $X$ and $Y$ remain the same and that the only difference in the joint distributions arises from the marginal distribution of the covariates – that is $P_s(Y|X) = P_t(Y|X)$ and $P_s(X) \neq P_t(X)$ – then the problem is known as *covariate shift* or *sample selection bias* (Shimodaira 2000; Huang et al. 2006). This scenario applies whenever there is a bias in the data selection procedure. For example, consider two different telescopes of which one is equipped with a sensor with higher sensitivity than the other. The data captured by the more sensitive telescope will be more biased towards dimmer objects than the one captured by the less sensitive telescope. However, the characteristics of the celestial objects do not change, that is, $P(Y|X)$ is the same. Figure 1 shows covariate shift between a sample of the EROS and HiTS datasets when looking at the mean magnitude and the Psi_CS feature from the FATS package (Nun et al. 2015).

### 2.2. Target, Conditional and Generalized Target Shift

Zhang et al. (2013) identify three distinct scenarios that arise if we assume that $P_s(Y|X) \neq P_t(Y|X)$. By virtue of Bayes' theorem, this difference is produced by the marginals $P(Y)$ or the conditionals $P(X|Y)$ being different, or both.

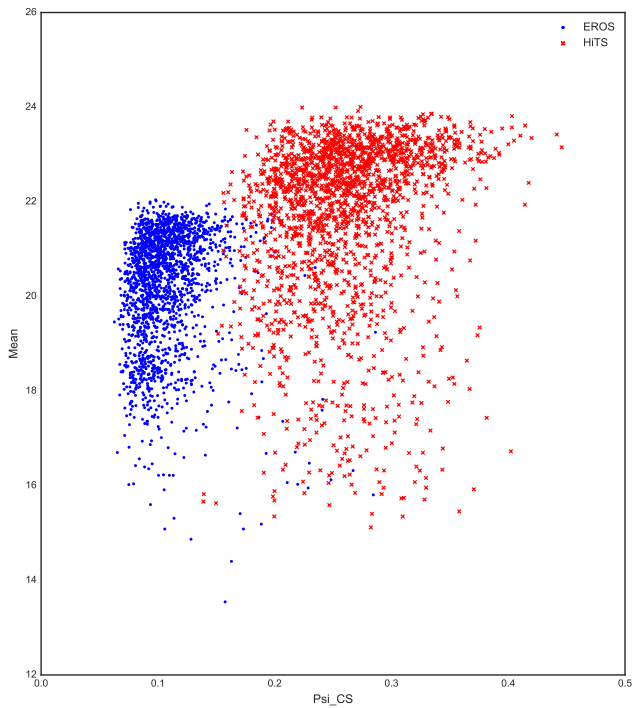

Fig. 1.— Covariate shift between EROS and HiTS datasets. The HiTS survey is more biased toward dimmer objects than EROS.

If the marginal distributions of the classes change and the conditional distributions of the features given the classes stay the same – that is $P_s(Y) \neq P_t(Y)$ while $P_s(X|Y) = P_t(X|Y)$ – then the problem is known as *target shift* (TarS) (Zhang et al. 2013), *class imbalance* (Patel et al. 2015) or *prior probability shift* (Quionero-Candela et al. 2009). This scenario arises whenever a class is more present in one domain than in the other. For example, in the problem of medical diagnosis prediction, one is interested in predicting diseases given symptoms. Disease prevalence varies across geographical locations, as such some diseases that are common in tropical regions will be rare in areas close to the poles and class distributions will be different (but the probability of the symptoms given the disease remains constant). In astronomy, if the sensitivity of the telescope changes significantly, then we see objects such as distant galaxies that are not in the other dataset.

Conversely, the problem is known as *conditional shift* (ConS) (Zhang et al. 2013) if the conditional distribution of the features given the classes changes and the marginal distribution of the classes stays the same – that is $P_s(Y) = P_t(Y)$ and $P_s(X|Y) \neq P_t(X|Y)$. ConS appears when the causal relationship of one or all the classes in relation to the features. For example, some diseases manifest symptoms differently depending on the patient's gender. The probability of having nausea given that the patient is suffering a heart disease will be higher if female patients are being diagnosed. In astronomy, variability may appear in some part of the electromagnetic spectrum. For example, a star may be variable in the ultraviolet, but not in the optical spectrum.

Finally, if both distributions change, meaning that $P_s(Y) \neq P_t(Y)$ and $P_s(X|Y) \neq P_t(X|Y)$, the problem is known as *generalized target shift* (GeTarS) (Zhang et al. 2013). This situation arises when both TarS and ConS are present.

Our research focuses on the domain adaptation problem under covariate and generalized target shift. Therefore, we do not assume that any of the marginal probability distributions are the same. Our goal is to find a transformation from the source feature space to the feature space of the target domain, in order to adapt the source labeled instances to a representation suitable for training a classifier that performs well in the target domain.

## 3. RELATED WORK

Domain adaptation has been studied extensively in the contexts of natural language processing (NLP) (Foster et al. 2010; Blitzer et al. 2006) and computer vision (Gong et al. 2012; Gopalan et al. 2011; Patel et al. 2015).

A popular approach for domain adaptation is known as *instance weighting* or *importance reweighting* (Shimodaira 2000; Foster et al. 2010). In instance weighting, the terms of the loss function corresponding to each sample are weighted by the relative density $\frac{P_t(x,y)}{P_s(x,y)}$, which effectively minimizes the expected loss of the model over the target distribution (Jiang 2008). However, it is generally not possible to calculate this value and the support of the source distribution must be contained in that of the target distribution for this to work in practice. In the covariate shift scenario, this weight can be simplified to $\frac{P_t(x)}{P_s(x)}$ (Shimodaira 2000). Under target shift, on the other hand, the weighting term is $\frac{P_t(y)}{P_s(y)}$. See Patel et al. (2015) for a more thorough explanation.

Daumé III (2009) proposes a feature augmentation framework in which features from both source and target domains are mapped into a feature space triple size of the original, which captures the feature similarities and particularities between source and target domains. In Daumé III's approach, given an input vector $x \in \mathbb{R}^n$, two mapping functions $\Phi_s(x) = [x, x, \mathbf{0}]$ and $\Phi_t = [x, \mathbf{0}, x]$ are defined for the source and target datasets, respectively. Here $\mathbf{0} = [0, 0, ..., 0] \in \mathbb{R}^n$ is the zero vector. In this way the enhanced feature space contains a general version of the data (the first third of the vector where data from both domains appear) and a version of the data specific to one of the domains (the second third of the vector for the source domain and the last third of the vector for the target domain). The classifier is then expected to learn the adaptation by, for example, assigning weights to each version of the data depending on how well it generalizes between the domains.

Other approaches are based on the idea of learning new feature representations that are domain invariant. Gopalan et al. (2011) developed a method motivated by incremental learning in which the adaptation is performed by gradually transitioning from one domain to the other. This is done by treating each domain as a point in the Grassmann manifold and sampling points along the geodesic path between them to obtain a description of the underlying domain shift. Gong et al. (2012) go further and integrate over an infinite number of subspaces using a kernel-based method.

Chan & Ng (2005) use expectation-maximization (EM) to estimate the class densities under the TarS setting by assuming the that the distribution of the features given the labels stay constant and applying the iterative procedure of the EM algorithm.

Kulis et al. (2011) introduce a method for finding nonlinear transformations between domains by learning in the kernel space.

Our method is similar to the *location-scale generalized target shift* (LS-GeTarS) transformation proposed by Zhang et al. (2013). In LS-GeTarS, a transformation from $P_s(X|Y)$ to $P_t(X|Y)$ is modeled as a translation and a scaling of the data given by $x^{\text{new}} = x \cdot W + B$, where $W$ is the scaling matrix and $B$ the translation. Instead of working directly with the marginal and conditional distributions they use their kernel mean embedding. A kernel mean embedding is a representation of a probability distribution as a point in a Reproducing Kernel Hilbert Space. In this manner, they do not need to assume a certain distribution, but minimize the loss using the kernel embedding and the algebraic operations it supports. The importance weight $\frac{P_t(Y)}{P_s(Y)}$ is estimated along the transformation parameters.

There are some other proposals that also use Bayesian transfer learning. Gönen & Margolin (2014) present a multi-task learning framework in which they apply kernel-based dimensionality reduction and use task-specific projection matrices to jointly find a common subspace. They define a different transformation of the data for each task, each of which is modeled as a projection matrix. The classifier is also part of the probabilistic model, and they do inference on the transformations and the classifier at the same time. Our work differs in two substantial ways: (1) we find a transformation from the feature space of one "task" (in our case of one survey) to the space of another one, while this method creates a new common space that is different from the original space of all the tasks; and (2) the specific classification task is not part of our model, only the transformation. This means that our method can be used with any classifier and that if we change the classifier we do not need to do inference again to find the transformation.

Another Bayesian method is proposed by Finkel & Manning (2009), who present a hierarchical Bayesian framework for multiple domain adaptation. For each domain, there is an arbitrary probabilistic model for which a normally distributed prior is put on its parameters. In the next level of the hierarchy, another normally distributed prior is added to the domain specific parameter priors. This hierarchy can be extended further for an arbitrary number of levels, reflecting related super-domains, super-super-domains, and so on.

## 4. BACKGROUND THEORY

### 4.1. *K-means*

The K-means algorithm (MacQueen et al. 1967) is a method for partitioning a set of points into a given number of clusters. Each cluster is defined in terms of a centroid and each point is assigned to the cluster of the closest centroid. Note that the clustering is completely defined in terms of the centroids. The learning algorithm initializes the centroids of a given number $K$ of clusters at random and then proceeds iteratively through two stages. In the first stage, each data instance is mapped to the cluster with the closest centroid. Then, in the second stage, each cluster centroid is re-calculated as the mean of all the points mapped to them. These steps are repeated until convergence (i.e. when no or very small variations in the centroid coordinates exist between two iterations). Figure 2 shows two iterations of an example execution of K-means.

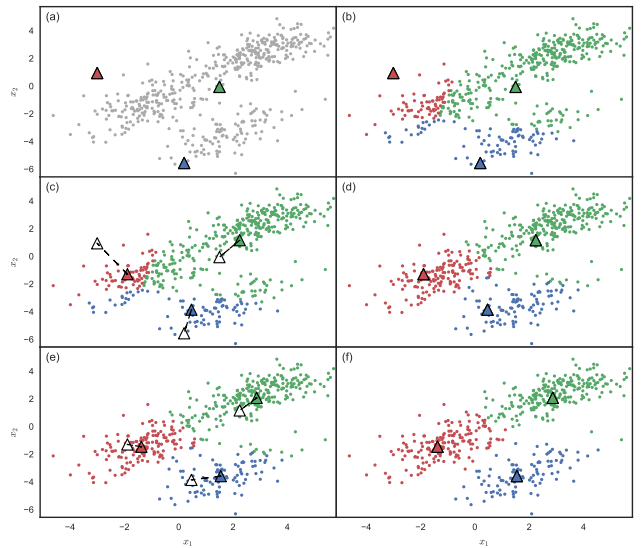

FIG. 2.— K-means example in 2D space. (a) Cluster centroids are initialized at random positions. (b) First step: each data point is mapped to the closest centroid. (c) Second step: the centroids are moved to the mean of each cluster's mapped points. (d, e, f) The steps are repeated iteratively until convergence.

#### 4.1.1. *Graphical Models and Plate Notation*

Probabilistic graphical models are a framework for conveniently representing and manipulating joint probability distributions – an thus Bayesian machine learning models – that draws from probability theory and graph theory. By using a graph data structure, graphical models leverage the wealth of representations and algorithms from computer science to allow for representation, learning and inference of otherwise unmanageable probability distributions.

In a graphical model, random variables are represented as nodes in a graph, while dependence relationships are encoded as graph edges. In this work, we are interested in a form of directed graphical models known as *Bayesian networks* (Koller & Friedman 2009), which encode a set of conditional independencies in a joint probability distribution. In a Bayesian network, a directed edge from variable $X$ to variable $Y$ indicates that variable $Y$ directly depends on variable $X$. The network satisfies the *local independency assumption*, which holds that every node $X_i \in \{X_1, ..., X_n\}$ of an $n$ node graph is conditionally independent of its non-descendants given its parents (Koller & Friedman 2009):

$$(X_i \perp \text{NonDescendants}_{X_i} \mid \text{Parents}_{X_i}) \quad (1)$$
$$\forall i \in 1, \ldots, n$$

In this work, we represent graphical models using the

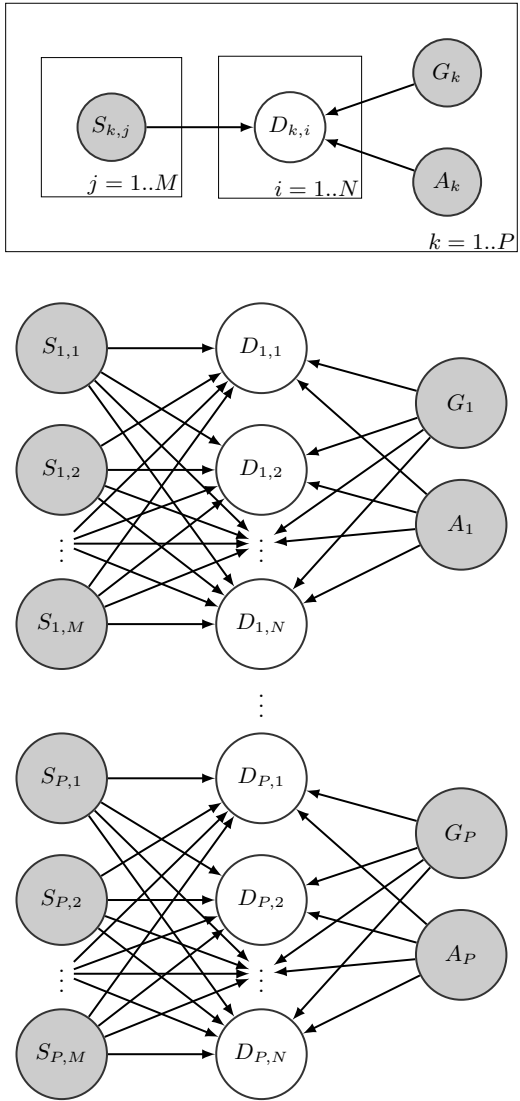

FIG. 3.— Example of a graphical model in plate notation. The top panel shows the model in plate notation. Indexed variables are displayed once. The bottom panel shows the same model using no plates. Each variable is explicitly displayed.

*plate notation* (Koller & Friedman 2009). Identically distributed variables that are repeated many times are enclosed by a rectangle or plate, capturing the notion that they correspond to a "stack" of identical variables. Variables in a plate are indexed and repeated according to the indication on the lower right corner of the plate. Edges connecting two nodes in the same plate connect variables with the same index. Edges connecting nodes inside a plate with nodes outside of it connect the outer variable with all the instances of the repeated variable. Edges from one plate to a different plate connect all instances in one plate with all the instances in the second plate. In this way, models can be represented in a more compact way by plotting each variable instance once. Consider a model for medical diagnosis with $P$ patients in which the presence of each of $N$ diseases $D_{k,1}, \ldots, D_{k,N}$ depends on the manifestation on patient $k$ of $M$ symptoms $S_{k,1}, \ldots, S_{k,M}$ and the gender $G_k$ and age $A_k$ of the patient. Figure 3 shows a graphical representation for this model, both using plates and by explicitly displaying all variables.

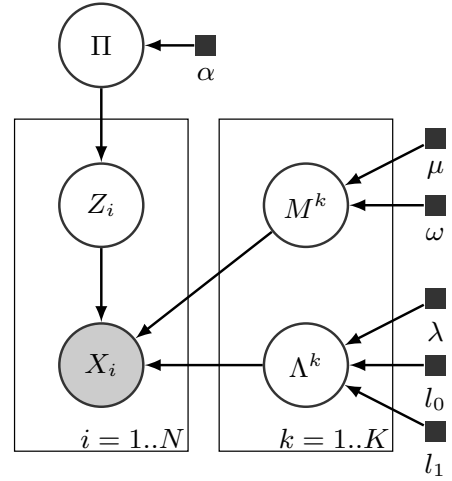

FIG. 4.— Gaussian mixture model in plate notation. Each data point $X$ is generated by the component indicated by $Z$. Each of the $K$ components has mean vector and precision matrix priors $M^k$ and $\Lambda^k$, respectively. Black squares indicate prior hyperparameters.

### 4.2. *Mixture Models*

A mixture model is a convex combination (or superposition) of probability distributions (i.e. the relative weights of the distributions sum to one) (Bishop 2006). Suppose we have $K$ Gaussian distributions generating a set of data. We call each distribution a component. Let $Z$ be a binary vector indicating which component generated a certain value of $X$, and give it a one-hot encoding representation where $Z^k$ is equal to 1 if $X$ was generated by component $k$ and 0 otherwise (e.g. $Z = [0\ 1\ 0\ 0]$ when the value is generated by component 2). Let $\Pi^k$ be the mixture coefficients. Then we can write the marginal distribution of $Z$:

$$P(Z) = \prod_{k=1}^{K} (\Pi^k)^{Z^k} \quad (2)$$

And the conditional distribution of $X$ given it was generated by component $k$ as:

$$P(X|Z) = \prod_{k=1}^{K} \mathcal{N}(X|M^k, \Sigma^k)^{Z^k} \quad (3)$$

Where $\mathcal{N}(X|M^k, \Lambda^k)$ is the probability density function for the multivariate Gaussian distribution with mean vector $M^k$ and covariance matrix $\Sigma^k$. Unfortunately, using covariance matrices is inefficient, as it involves expensive matrix inversion operations when computing the likelihood. For this reason, we use the inverse covariance matrix, also known as precision matrix, $\Lambda^k = (\Sigma^k)^{-1}$:

$$P(X|Z) = \prod_{k=1}^{K} \mathcal{N}(X|M^k, \Lambda^k)^{Z^k} \quad (4)$$

We use equations (2) and (4) to obtain the mixture's density by marginalizing over $Z$:

$$\begin{aligned} P(X) &= \sum_Z P(X|Z)P(Z) \\ &= \sum_Z \prod_{k=1}^K \mathcal{N}(X|M^k, \Sigma^k)^{Z^k} \prod_{k=1}^K (\Pi^k)^{Z^k} \\ &= \sum_Z \prod_{k=1}^K \left(\Pi^k \, \mathcal{N}(X|M^k, \Sigma^k)\right)^{Z^k} \\ &= \Pi^1 \, \mathcal{N}(X|M^1, \Sigma^1) + ... + \Pi^K \, \mathcal{N}(X|M^K, \Sigma^K) \\ &= \sum_{k=1}^K \Pi^k \mathcal{N}(X|M^k, \Lambda^k) \end{aligned}$$
(5)

Thus, we can interpret each $\Pi^k$ as the prior probability of assigning a sample $X$ to component $k$ (Bishop 2006).

We can derive the corresponding posterior probability $\Gamma$ using Bayes' theorem:

$$\begin{aligned} \Gamma^k(X) &= P(Z^k = 1|X) \\ &= \frac{\Pi^k \mathcal{N}(X|M^k, \Lambda^k)}{\sum_{l=1}^K \Pi^l \mathcal{N}(X|M^l, \Lambda^l)} \end{aligned}$$
(6)

$\Gamma^k$ is referred to as the responsibility of component $k$ and it represents how strongly component $k$ contributed to generating sample $X$ (Bishop 2006).

Figure 4 shows a Gaussian Mixture model in plate notation.

### 4.3. Precision Matrix Modeling

As we are developing a Bayesian model, we assign priors to the mixture component's parameters. For the components' means a Gaussian prior is commonly used. As for the precisions, the Wishart distribution is a popular choice due to its conjugacy to the multivariate Gaussian distribution when a dependency to the mean is introduced. However, as Barnard et al. (2000) point out, when specifying a prior it is more convenient to work in terms of standard deviation and correlation. For this purpose, they suggest a separation strategy, decomposing a covariance matrix $\Sigma$ into a standard deviation vector $\sigma$ and a correlation matrix $C$:

$$\Sigma = \text{diag}(\sigma) \, C \, \text{diag}(\sigma) \quad (7)$$

Here $\text{diag}(v)$ represents the square diagonal matrix whose main diagonal contains the elements in vector $v$. This provides the advantage that we can express our prior knowledge of the standard deviation and correlation separately on the original scale of the standard deviation (Barnard et al. 2000).

A random value for $\sigma$ can be generated using any continuous distribution. We generate the correlation matrix $C$ using the method proposed by Lewandowski et al. (2009), from which we can get a random correlation matrix $C$ of any given dimension with density proportional to $|C|^{\lambda-1}$ for a shape parameter $\lambda > 1$. We will refer to this distribution over the space of correlation matrices as LKJ. In section 5.1 we explain how we use this principle to model the precision matrix priors for the mixture components.

### 4.4. N-Dimensional Rotations

Rotations in 2D and 3D space are commonly understood as rotations about an axis by a certain angle. Duffin & Barrett (1994) argue that it is better to think about them as occurring in a plane: the plane perpendicular to the axis of rotation in 3D, and the only plane in 2D. They generalize the concept to rotation in n-dimensional space in principal planes formed by two coordinate axes.

The rotation matrix for the rotation of axis $X_a$ in the direction of axis $X_b$ by an angle of $\theta$ is as follows (Duffin & Barrett 1994):

$$R_{ab}(\theta) = \left\{ r_{ij} \left| \begin{array}{ll} r_{ii} = 1 & i \neq a, i \neq b \\ r_{aa} = \cos\theta & \\ r_{bb} = \cos\theta & \\ r_{ab} = -\sin\theta & \\ r_{ba} = \sin\theta & \\ r_{ij} = 0 & \text{elsewhere} \end{array} \right. \right\} \quad (8)$$

An arbitrary rotation in n-dimensional space can then be specified as the composition of rotations in the $\binom{d}{2}$ principal planes.

## 5. METHOD DESCRIPTION

We propose a probabilistic model based on the GMM to describe the $P_s(X, Y)$ and $P_t(X, Y)$ distributions. We model mixture weights, component mean vectors and precision matrices for the source distribution in the usual manner. However, each of the target distribution mixture parameters is modeled as a separate transformation of the respective parameter of the source mixture: each target mean vector is equal to the respective source mean vector plus a translation and each target precision matrix is equal to a scaling and a rotation of the respective source precision matrix. Note that each component needs not undergo the same transformation as the others, since there are separate transformation variables for each one. Here, we are making the assumption that we can capture the domain shift between the datasets as a mixture of transformations in the subspaces defined by each multivariate Gaussian. That is, we propose a model that describes a mixture of Gaussians over the source dataset and a linear transformation for each of its components, which result in a transformed mixture of Gaussians over the target dataset. Note that since the target mixture is fully determined by the source mixture and the transformation, the mapping between the corresponding components is given implicitly – the target component that corresponds to a source component is simply the one resulting from applying the component's transformation. Furthermore, we assume that the training set suffers the same shift between the domains than the unlabeled dataset and that by inferring it from the latter we will be able to correct it for the former as well. The latter implies that we assume that there is no significant unrepresented population in the unlabeled data.

Our method comprises 5 steps, as shown in figure 5: (1) the model is set-up using the unlabeled data from the source and target domains (the detailed model specification follows in section 5.1). An optional step of randomly sampling from the datasets may be performed here, depending on the amount of data and computational resources available. (2) The mixture and transformation parameters variables are sampled using the Metropolis Hastings MCMC method. (3) The samples are used to make a point estimate of the parameters. Steps 2 and 3 are explained in section 5.2. (4) The estimated parameters are used to apply the modeled transformation to the training set that is available for the source domain, in order to correct the shift. The transformation is explained in detail in section 5.3. (5) Using the transformed training data, a classifier expected to perform well on the target domain is trained. (*) In our experiments, presented in section 7, we perform an additional step of testing on a target domain labeled dataset in order to assess the method's performance. This dataset is not used at any moment in the previous steps.

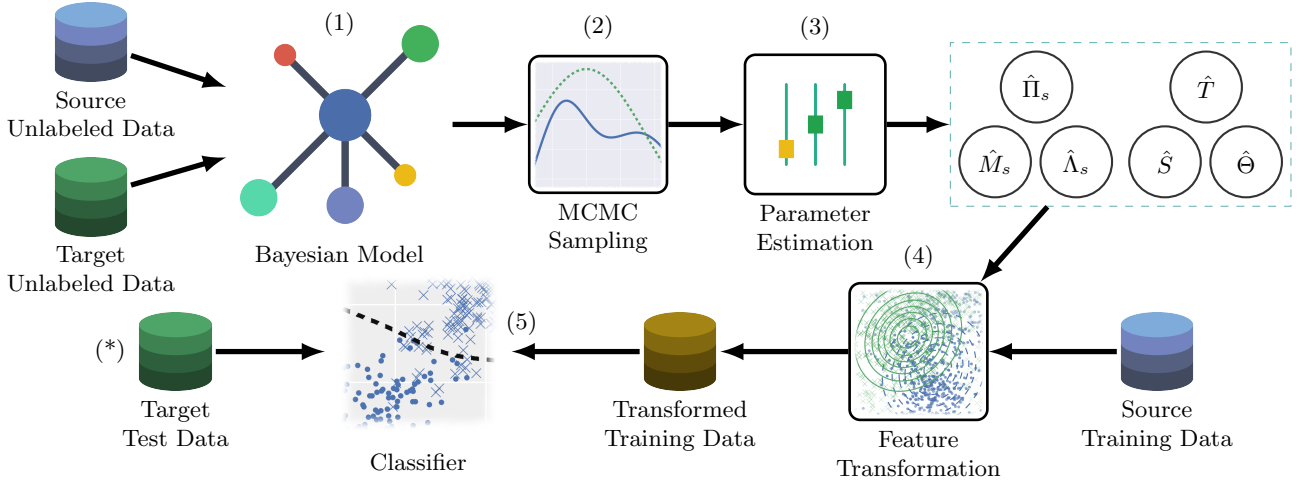

FIG. 5.— Domain adaptation method overview. (1) The probabilistic model is constructed according to the specification and the unlabeled input data. (2) MCMC techniques are used to sample from the posterior distributions of the transformation parameters. (3) The transformation parameters are estimated by averaging the samples. (4) The transformation is applied to the source training data according to the estimated parameters. (5) A classifier is trained on the transformed input data from the source domain. (*) In our experiments, the classifier is tested on labeled data from the target domain to assess performance.

### 5.1. Model Specification

First, we specify the mixtures that represent the source and target datasets. Let $X_s^i$ and $X_t^j$ be random variables for the $i$-eth and $j$-eth unlabeled sample in the source and target datasets, respectively, and let $d$ be the dimensionality of the data. Let $Z_s^i$ and $Z_t^j$ denote the component assignments for source sample $i$ and target sample $j$, $M_s^k$ and $M_t^k$ the mean for component $k \in 1..K$ and $\Lambda_s^k$ and $\Lambda_t^k$ the precision for component $k$ in the source and target domains, respectively. Let $\Pi_s$ and $\Pi_t$ be the priors for the component weights.

The following distributional assumptions are made:

$$\Pi_s \sim \mathcal{D}(\alpha) \qquad\qquad H \sim \mathcal{D}(\eta)$$
$$Z_s^i \sim \mathcal{C}(\Pi_s) \quad \forall i \qquad Z_t^j \sim \mathcal{C}(\Pi_t) \quad \forall j$$
$$M_s^k \sim \mathcal{N}(\mu, \omega) \quad \forall k$$
$$X_s^i \sim \mathcal{N}(M_s^{Z_s^i}, \Lambda_s^{Z_s^i}) \quad \forall i \qquad X_t^j \sim \mathcal{N}(M_t^{Z_t^j}, \Lambda_t^{Z_t^j}) \quad \forall j$$

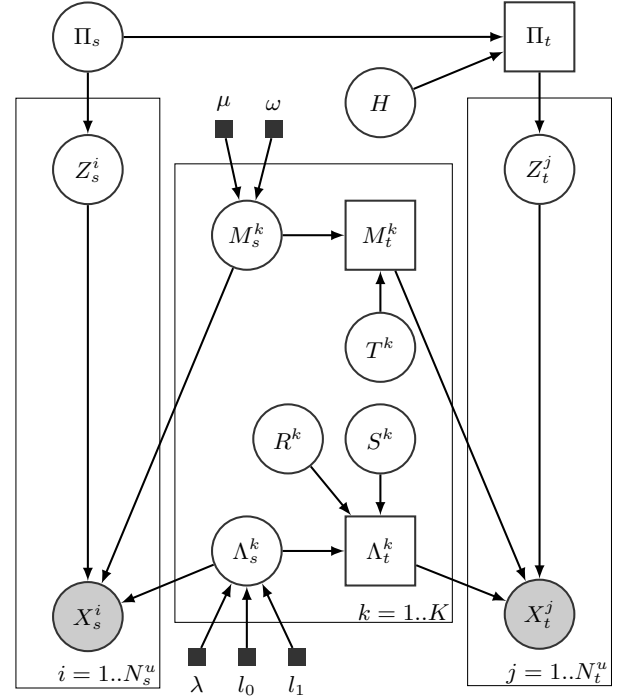

FIG. 6.— The proposed model in plate notation. Random variables are shown in circles. Variables derived deterministically from other variables are shown in squares. Prior hyperparameters are shown as small black squares. Observed variables are shaded in gray. Some hyperparameters are omitted for clarity.

Where $\mathcal{D}(\alpha)$ denotes the Dirichlet distribution with concentration parameter vector $\alpha$ and dimension $K$, $\mathcal{C}(\pi)$ represents the categorical distribution with event probability vector $\pi$ of dimension $K$, and $\mathcal{N}(\mu, \Lambda)$ represents the normal distribution of dimension $d$ with mean vector $\mu$ and precision matrix $\Lambda$.

The target component weights are determined by the source component weights and the Dirichlet distributed variable $H$, which scales each weight like so:

$$\Pi_t^k = \Pi_s^k H^k / \sum_{l=1}^{K} \Pi_s^l H^l \quad \forall k \qquad (9)$$

Higher values for the hyperparameter $\eta$ will favor small differences in component weight between domains.

As explained in section 4.4, the source domain components' precision matrices $\Lambda_s^k$ are generated using the separation strategy defined in equation (7). Each resulting covariance matrix is then inverted to yield the corresponding precision matrix:

$$L^k \sim \mathcal{LKJ}(\lambda) \quad \forall k \qquad \Sigma^k \sim \mathcal{U}(l_0, l_1) \quad \forall k$$
$$\Lambda_s^k = \left(\operatorname{diag}(\Sigma^k) L^k \operatorname{diag}(\Sigma^k)\right)^{-1} \quad \forall k$$

Where $\mathcal{U}(l_0, l1)$ denotes the uniform distribution of dimension $d$ with minimum value $l_0$ and maximum value $l_1$, and $\mathcal{LKJ}(\lambda)$ denotes the LKJ distribution of dimension $d$ with shape parameter $\lambda$.

Second, we introduce random variables for the component transformations. Each component suffers a translation of its mean vector, and a rotation and a scaling of its precision matrix. The translation of component's $k$ mean, $T^k$, and the resulting target mean vectors $M_t^k$ are as follows:

$$T^k \sim \mathcal{N}(0, \kappa I) \quad \forall k$$
$$M_t^k = M_s^k + T^k \quad \forall k$$

Where $0$ represents the zero vector of dimension $d$ and $I$ represents the $d \times d$ identity matrix. The $\kappa$ hyperparameter specifies the a priori belief about the magnitude of the translations, so that smaller $\kappa$ values will favor larger translations.

Each precision matrix rotation is modeled as a $\binom{d}{2}$-dimensional vector of angles $\Theta^k$, representing the rotation of each principal plane. The precision matrices scalings are modeled as factors multiplying each dimension centered at the component's mean:

$$\Phi^k \sim \mathcal{B}^{\binom{d}{2}}(\beta, \beta) \quad \forall k \qquad S^k \sim \mathcal{B}^d(\epsilon, \epsilon) + 0.5 \quad \forall k$$
$$\Theta^k = (2\Phi^k - 1)\rho$$

The notation $\mathcal{B}^d(\alpha, \beta)$ corresponds to the beta distribution of dimension $d$ with shape parameters $\alpha$ and $\beta$. We abuse the notation $\mathcal{B}^d(\alpha, \beta) + \delta$ to represent the same beta distribution with its range offset by $\delta$, resulting in a support in the range $[\delta, 1.0 + \delta]$. We let the rotation in each principal plane be in the interval $[-\rho, \rho]$. In order to do so, we draw from each $\Phi^k$ prior $\binom{d}{2}$ values between 0 and 1 and use them to interpolate between the rotation limits and get $\Theta^k$, the angles of rotation. The hyperparameter $\beta$ represents the a priori belief about the magnitude of the rotations. Larger values of $\beta$ mean a more tight distribution around 0.5, which equals to a null rotation. Then, for each of the $\binom{d}{2}$ main planes of rotation we use equation 8 to build a rotation matrix and compose them like so:

$$R^k = \overrightarrow{\prod}_{a=1}^{d-1} \overrightarrow{\prod}_{b=a+1}^{d} R_{ab}(\Theta_l^k) \quad \forall k \qquad (10)$$

$$l = (a-1)(d - a/2) + b - a \qquad (11)$$

Where $\overrightarrow{\prod}$ denotes aggregated left side matrix multiplication so that equation (10) expands to:

$$\begin{aligned} R^k =& R_{(d-1)\,d}\left(\Theta_{\binom{d}{2}}^k\right) \ldots R_{2d}(\Theta_{2d-3}^k) \ldots \\ & \ldots R_{24}(\Theta_{d+1}^k)\, R_{23}(\Theta_d^k)\, R_{1d}(\Theta_{d-1}^k) \ldots \\ & \ldots R_{13}(\Theta_2^k)\, R_{12}(\Theta_1^k) \qquad \forall k \end{aligned} \qquad (12)$$

The scaling factors for each dimension are allowed in the range $[0.5, 1.5]$. The $\epsilon$ determines how tightly around a 1 scaling factor the distribution will be.

We get the precision matrix that would result from scaling the data in each component $k$ by $S^k$ and then rotating it according to $R^k$ as:

$$\Lambda_t^k = R^k (S^k)^{-1} \Lambda_s^k (S^k)^{-1} (R^k)^{-1} \quad \forall k \qquad (13)$$

### 5.2. Parameter Estimation

In order to apply the transformation, we first estimate the $T^k$, $S^k$ and $\Theta^k$ transformation parameters, and the $\Pi_s$, $\Pi_t$, $M_s^k$, $\Lambda_s^k$ model parameters by sampling from their posterior distributions using the Gibbs MCMC sampler. To accelerate convergence, we run KMeans on the source dataset to find centroids for the source components and initialize the mean vectors to their values. We step through MCMC iterations until the standard deviation of the samples is below a certain threshold. We then use the mean point estimate of the samples as the parameter values.

### 5.3. Feature Transformation

Let $\hat{\Pi}_s$ and $\hat{\Pi}_t$ be the estimates for the source and target component weights $\Pi_s$ and $\Pi_t$, respectively. Let the $K \times d$ matrix $\hat{T}$ contain the estimate of the translation $T^k$ of each component as rows, the $K \times d$ matrix $\hat{S}$ contain the estimates of the component scalings $S^k$ as rows, and the $K \times \binom{d}{2}$ matrix $\hat{\Theta}$ contain the estimate for the rotation angles of the $\binom{d}{2}$ principal planes of each component as rows. Similarly, let $\hat{M}_s^k$ and $\hat{\Lambda}_s^k$ be the estimates for the mean vector and the precision matrix for each component $k$, respectively.

We then apply a transformation $\Psi$ to the source domain training set $D_s^l$ in order to obtain a labeled dataset $D_\star^l = \{(\Psi(x_i^{sl}), y_i^{sl})\}_{i=1}^{N_s^l}$ suitable for training a classifier for the target domain. Let $X$ be a $N_s^l \times d$ matrix containing the source training samples, such that $X_i = x_i^{sl}$ for $i = 1, ..., N_s^l$.

We compute the $N_s^l \times K$ matrix $W$ containing the component transformation weights for each instance given by equation 6:

$$W_{ik} = \Gamma^k(X_i) \quad \forall i, \forall k \qquad (14)$$

The translation for each instance is given by the weighted average of the component responsibilities and the component translations:

$$\Delta = W\hat{T} \quad (15)$$

Where $\Delta$ is a $N_s^l \times d$ matrix containing the translation for each instance.

Scaling proportional to component responsibility is applied by computing the translation $\Xi_i$ that results from scaling centered under each component:

$$\Xi_i = \sum_{k=1}^{K} W_{ik}\left(\text{diag}(\hat{S}^k) - I\right)(X_i - \hat{M}_s^k) \quad \forall i \quad (16)$$

Where $I$ is the $d \times d$ identity matrix and $\text{diag}(v)$ is the square diagonal matrix whose main diagonal contains the elements in vector $v$.

Finally, we rotate the data with respect to each component center. First we compose the rotation matrices for each instance and component similarly as in equation 10, but using weighted transformation angles:

$$\hat{R}_i^k = \overrightarrow{\prod_{a=1}^{d-1}} \overrightarrow{\prod_{b=a+1}^{d}} R_{ab}(W_{ik}\hat{\Theta}_l^k) \quad \forall i, \forall k \quad (17)$$

With $l$ as given by equation 11.

Then the transformation $\Psi(X_i) = X^\star$ is given by the following algorithm:

$X^\star = X_i + \Xi_i$
**for** $k = 1 \to K$ **do**
$\quad X^\star = \hat{R}_i^k(X^\star - \hat{M}_s^k) + \hat{M}_s^k$
**end for**
$X^\star = X^\star + \Delta_i$

Which applies the offset produced by the scalings, rotates the data according to each component and finally applies the weighted translation.

## 6. IMPLEMENTATION

For our implementation we used PyMC3 (Salvatier et al. 2016) for the model specification and MCMC sampling. For data manipulation, linear algebra, and statistical and numerical computation we used NumPy (Walt et al. 2011), SciPy (Walt et al. 2011), Pandas (McKinney 2010) and Theano (Al-Rfou et al. 2016). The plots shown in this paper were generated using Matplotlib (Hunter 2007). We also used the SVM, RF and metrics implementations of Scikit-learn (Pedregosa et al. 2011).

The implementation of our model and the data used in our experiments is available as a Python package at https://goo.gl/EPgnkk.

## 7. EXPERIMENTAL RESULTS

### 7.1. *Methodology*

We transfer the training knowledge from the source to the target catalog by performing the steps illustrated in figure 5:

1. Build the Bayesian model with the source catalog and target catalog unlabeled datasets.

TABLE 1
F1 SCORES FOR SIMULATED CONS AND GETARS EXPERIMENTS.

| Classifier | ConS | | GeTarS | |
|---|---|---|---|---|
| | Original | Transformed | Original | Transformed |
| SVM | 88% | 95% | 87% | 93% |
| RF | 84% | 94% | 85% | 91% |

2. Perform MCMC sampling from the posterior distributions of the transformation parameters until convergence.

3. Estimate the transformation parameters using the mean of the samples.

4. Take the source catalog's training set and transform it using the parameters obtained in the previous step.

5. Train a classifier using the adapted training set.

We then measure the performance of our method by testing the classifier on a labeled dataset from the target catalog left out for this purpose.

The classifiers used in our experiments are the Radial Basis Function (RBF) kernel Support Vector Machine (SVM) (Boser et al. 1992) and the Random Forest (RF) (Breiman 2001) classifier.

### 7.2. *Simulations*

We generated datasets and simulated domain shifts of different nature in order to study the performance and behavior of our model under conditional shift and generalized target shift.

First, we simulated covariate and conditional shift. The left two panels in figure 7 show a simulated dataset generated using two multivariate Gaussians. In the top panel, the target dataset, shown in green, was generated by translating, scaling and rotating the source dataset distribution, shown in blue. The components of the fitted model are represented as level curves. The bottom panel shows the transformed source dataset in yellow, along with the original source dataset in blue.

Then, we simulated the GeTarS scenario by having a different class distribution between the source and target datasets. The two right panels in figure 7 show the generated datasets and the fitted model in the same manner.

Figure 8 shows how the decision boundary of a radial basis kernel (RBF) support vector machine (SVM) adapts to classify better in the target dataset when trained with the transformed data.

The classification F1 scores for both experiments are shown in table 1.

Note that even tough that each mixture component suffers a linear transformation, the overall transformation is not necessarily linear since the transformation under each component is combined with the others. For example, it is possible to rotate or scale some parts of the space while maintaining others relatively constant. This suggests that it is possible to capture more complex transformations, such as non-affine transformations where collinearity, line parallelism, convexity, length and area ratios, etc., are not preserved.

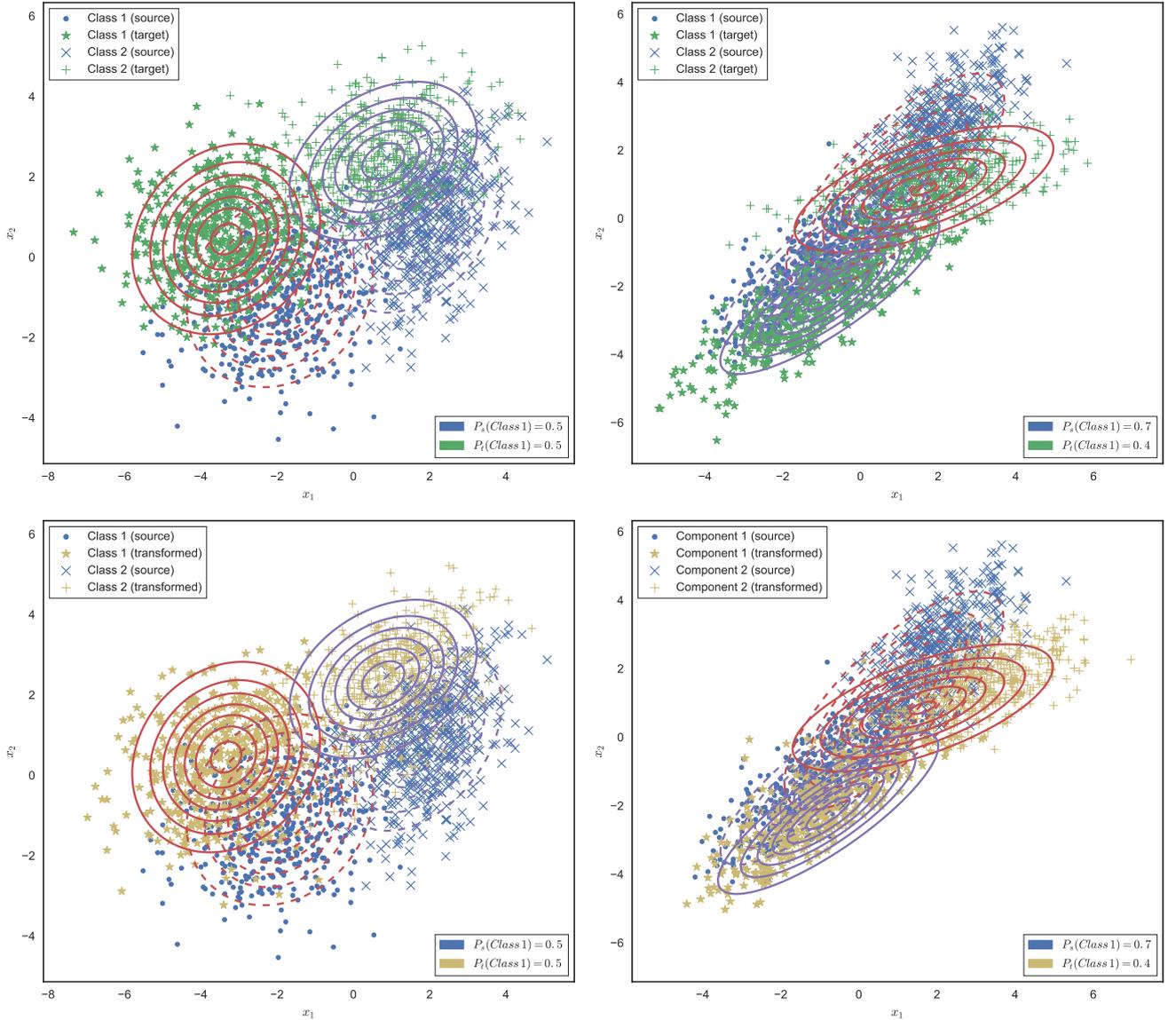

Fig. 7.— Model visualization on simulated data. Left two panels: simulation under ConS. Right two panels: simulation under GeTarS. Top panels show the fitted models. The dashed level curves represent the mixture components for the source dataset and the continuous lines show the components after applying the transformation found. the bottom panels show the transformation. The source dataset points are shown in blue along with their image after applying the transformation in yellow. The class distributions for each dataset are shown on the lower right corner.

### 7.3. Real Datasets

We apply our method to variable star classification using lightcurves from three different survey catalogs: the *Expérience pour la Recherche d'Objets Sombres II* (EROS) survey, the Massive Compact Halo Object (MACHO) survey and the High Cadence Transient Survey (HiTS). Sections 7.3.1, 7.3.2 and 7.3.3 contain a brief description of each survey, followed by a comparison in section 7.3.4.

We extract features from the lightcurves using the Feature Analysis for Time Series (FATS) Python package (Nun et al. 2015). In addition to the two top features according to the importance ranking presented by Nun et al. (2015), we selected the mean observed magnitude (herein referred to as "Mean") and the skewness of the observed magnitudes (herein referred to as "Skew"). We know that the mean magnitude is a proxy of the absolute magnitude for MACHO and EROS. Since these two surveys observe the Magellanic Clouds, the distance to the observed stars is approximately constant. We found that using 5 mixture components was enough to get reasonable results. The extracted features are described in table 2.

#### 7.3.1. The EROS Survey

The *Expérience pour la Recherche d'Objets Sombres II* (EROS-II or simply referred to as EROS in this paper) collaboration is an astronomical survey that started operation in 1990 at the European Southern Observatory at La Silla, Chile. Its main purpose was to search for microlensing events in the directions of the Magellanic Clouds, the Galactic Center and four areas within the Galactic Plane (Beaulieu et al. 1995). The EROS-II instrument was a 0.98 m diameter Ritchey-Chrétien telescope located at the European Southern Observatory in La Silla, Chile. It operated at $f/5$ with a 0.7°RA and 1.4°Dec field of view. The telescope featured a dichroic beam splitter that allowed for simultaneous observations in two wide pass-bands – a blue one and a red one. Perdereau (1998); Bauer et al. (1998); Observatory (2017).

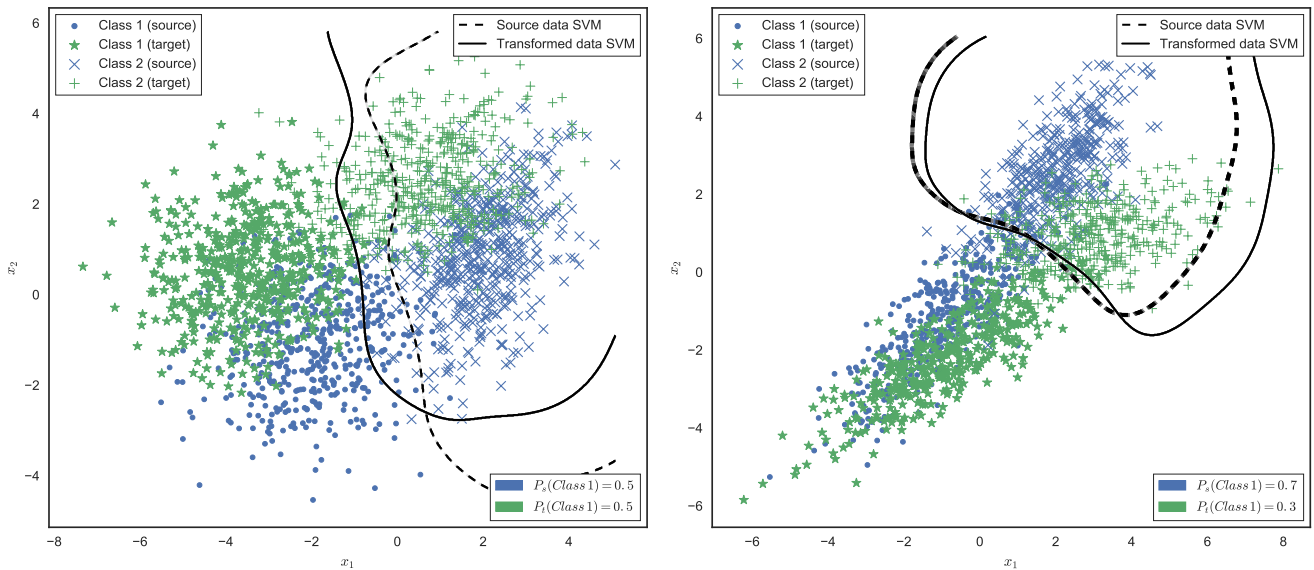

FIG. 8.— SVM classifier adaptation visualization. Decision boundaries for RBF kernel SVM's trained on the source training set and the transformed source training set. Left panel: simulation under ConS. Right panel: decision boundaries in a second simulation under GeTarS. The class distributions for each dataset are shown on the lower right corner.

### 7.3.2. *The MACHO Survey*

The Massive Compact Halo Object (MACHO) project is a gravitational microlensing survey whose main goal was to find massive compact halo objects in the Milky Way halo to assess their mass contribution (Alcock et al. 1997). Observations were made in the direction of the Large Magellanic Cloud (LMC), the Small Magellanic Cloud (SMC) and the Galactic Bulge. The MACHO project instrument was the 1.27 meter telescope at Mount Stromlo Observatory, Australia. It operated at $f/3.8$ with a 1°diameter field of view. A dichroic beamsplitter and filters allowed image capture in the 'red' (approx. 6,300 - 7,800 Å) and 'blue' bands (approx. 4,500 - 6,300 Å). Each image has a sky coverage of 0.72 x 0.72 degrees. The exposure times were of 300 seconds for the LMC, 600 seconds for the SMC and 150 seconds for the bulge (Alcock et al. 1997; Hart et al. 1996; Cook 1995).

In this paper we only consider the LMC data.

### 7.3.3. *The HiTS Survey*

The High Cadence Transient Survey (HiTS) first campaign started in 2013 with the objective of exploring transient and periodic objects with characteristic timescales between a few hours and days. This discovery survey uses high cadency data obtained from the Dark Energy Camera (DECam) mounted on a 4 m telescope at Cerro Tololo Interamerican Observatory (CTIO). The large *etendue* (product of collecting area and field of view) of the DECam allows the observation of apparent magnitudes as low as 24.5 mag. It operated at $f/2.7$ with a 2.2°field of view. (Förster et al. 2016; Flaugher 2006; Fukugita et al. 1996).

### 7.3.4. *Dataset Comparison*

Among the three surveys studied, MACHO and EROS are the most similar. They observed along two comparable bands, had an analogous limiting magnitude, and we used data from the same observed area for our experiments. However, as we can see from tables 5, 6, and 10, the classification performance drops significantly when training in one of these datasets and classifying in the other. F1 score drops from 85% to 60% when training in MACHO and classifying in EROS using Random Forest with four features, versus training and testing in MACHO. When training in EROS and classifying in MACHO, the drop is from 90% to 71%.

In contrast, EROS and MACHO are comparatively more dissimilar than HiTS. Some differences are: HiTS observed in four bands instead of two, it had a limiting magnitude around four points higher, a wider field of view, and it observed a different region of the sky. Table 3 shows a comparison of the three surveys and

TABLE 2
Features used in the experiments

|   | Name | Description |
|---|------|-------------|
| 1 | Color | Difference between the mean magnitude of observations from two different bands. |
| 2 | Mean | Mean magnitude. The arithmetic mean of all the lightcurve observations. |
| 3 | Psi_CS | Range of a cumulative sum applied to the phase-folded light curve (using the period estimated from the Lomb-Scargle method). |
| 4 | Skew | The skewness of the observed magnitudes. |

**Notes.** Names and descriptions are as in the FATS package. See Nun et al. (2015) for a detailed definition.

TABLE 3
Telescope and survey comparison

|  | EROS | MACHO | HiTS |
|---|---|---|---|
| Instrument | "MarLy" Ritchey-Chrétien Telescope | Great Melbourne Telescope (Renovation) | Víctor M. Blanco Telescope |
| Institution | European Southern Observatory | Australian National University | Cerro Tololo Interamerican Observatory |
| Location | La Silla, Chile | Mount Stromlo, Australia | Cerro Tololo, Chile |
| Altitude (masl) | 2,375 | 770 | 2,207 |
| Diameter (m) | 0.98 | 1.27 | 4 |
| Aperture | f/5 | f/3.8 | f/2.7 |
| Field of view | 0.7°(RA) 1.4°(Dec) | 1° | 2.2° |
| Bands (Å) | Blue (4,200 - 6,500) Red (6,500 - 9,000) | Blue (4,500 - 6,300) Red (6,300 - 7,800) | Blue-green (4,000 - 6,200) Red (4,850 - 7,650) Far-red (6,200 - 9,200) Near-infrared (7,900 - 10,000) |
| Limiting mag. | 20 | 20.5 | 24.5 |
| Observed area[a] | Magellanic Clouds | Magellanic Clouds | Southern Galactic Cap |

**Notes.**
[a] Of the data used in the experiments. See the survey description for the complete observation area.

TABLE 4
Dataset class composition

|  | Class | Description | # EROS | # MACHO | # HiTS |
|---|---|---|---|---|---|
| 1 | CEP | Cepheids. | 472 | 14 | 35 |
| 2 | EB | Eclipsing Binaries. | 12,061 | 207 | 15 |
| 3 | QSO | Quasars. | 217 | 55 | 2,309 |
| 4 | RRLYR | RR Lyrae. | 11,787 | 611 | 105 |
| 5 | LPV | Long Period Variables. | 1,468 | 217 | 0 |
|  |  | Total | 26,005 | 1,104 | 2,464 |

their instruments. Unsurprisingly, the classification performance drops dramatically when using HiTS as a training set for classifying in EROS or MACHO, and vice versa. When classifying in HiTS using Random Forest and four features, the F1 score drops from 94% to 8% when training in EROS, and to 11 % when training in MACHO. When using HiTS to train, the F1 score using Random Forest and four features drops from 85% to 3% classifying in EROS, and from 90% to 3% classifying in MACHO.

The datasets used in our experiments are also different in the amount of labeled data available. While our labeled dataset for EROS has more than 25,000 labeled stars, the MACHO and HiTS datasets have only about 1,000 and 2,5000, respectively. Moreover, the class representation is different in each dataset. Table 4 shows a description and the amount of instances for each class present in the labeled datasets of each survey.

### 7.3.5. Baseline Results

Tables 5, 6 and 7 present the per-class classification F1 scores obtained by crossvalidation in each dataset. These results serve as a baseline for the performance that can be achieved by both training and testing in a same dataset using the same features we transfer in our experiments.

### 7.3.6. 2D Experiment Visualization

To illustrate the functioning of the model, we first apply it to a two-dimensional space using the Mean and Color features (see table 2). We use EROS as the source dataset and MACHO as the target. Hence our goal is to classify MACHO instances using the transformed EROS training set. We use 10,000 unlabeled instances from each of the domains to fit our transformation model with 5 components. Figure 9 shows the transformation found for the four components with the highest weights. Note how each component "focuses" on a different region of the distribution and then transforms instances to "match' the target distribution region.

### 7.3.7. Further Experiments

We continue applying the method to an increasing number of features from table 2. We repeat the experiments using all possible dataset pairs. Table 8, 9 and 10 show the F1 scores for this experiment when transforming two, three and four features, respectively. The "Original" column displays the score obtained when training in the untransformed source dataset and testing on the target dataset. The "Transformed" column shows the score when training on the transformed source dataset and testing on the target dataset.

Tables 11 through 28 show the F1 scores for each class in each experiment when transforming two, three, and four features.

### 8. CONCLUSION

We present a method for survey invariant classification of variable stars by transforming feature representations

TABLE 5
Baseline F1 scores for variable star classification in EROS

|   | Class | Random Forest | | | Support Vector Machine | | |
|---|---|---|---|---|---|---|---|
|   |   | 2 Features | 3 Features | 4 Features | 2 Features | 3 Features | 4 Features |
| 1 | CEP | 53% | 72% | 79% | 32% | 16% | 72% |
| 2 | EB | 75% | 80% | 85% | 76% | 78% | 84% |
| 3 | QSO | 1% | 21% | 24% | 0% | 0% | 0% |
| 4 | RRLYR | 74% | 80% | 84% | 77% | 79% | 83% |
| 5 | LPV | 90% | 92% | 92% | 91% | 91% | 92% |
|   | Weighted Average | 74% | 80% | 85% | 76% | 77% | 83% |

TABLE 6
Baseline F1 scores for variable star classification in MACHO

|   | Class | Random Forest | | | Support Vector Machine | | |
|---|---|---|---|---|---|---|---|
|   |   | 2 Features | 3 Features | 4 Features | 2 Features | 3 Features | 4 Features |
| 1 | CEP | 52% | 61% | 69% | 13% | 13% | 80% |
| 2 | EB | 73% | 74% | 82% | 65% | 66% | 82% |
| 3 | QSO | 33% | 67% | 59% | 0% | 0% | 10% |
| 4 | RRLYR | 89% | 91% | 93% | 89% | 89% | 93% |
| 5 | LPV | 98% | 98% | 98% | 98% | 98% | 98% |
|   | Weighted Average | 84% | 88% | 90% | 81% | 81% | 87% |

TABLE 7
Baseline F1 scores for variable star classification in HiTS

|   | Class | Random Forest | | | Support Vector Machine | | |
|---|---|---|---|---|---|---|---|
|   |   | 2 Features | 3 Features | 4 Features | 2 Features | 3 Features | 4 Features |
| 1 | CEP | 35% | 41% | 35% | 10% | 10% | 33% |
| 2 | EB | 0% | 42% | 33% | 0% | 0% | 0% |
| 3 | QSO | 97% | 97% | 98% | 97% | 97% | 97% |
| 4 | RRLYR | 24% | 36% | 46% | 22% | 23% | 19% |
|   | Weighted Average | 92% | 94% | 94% | 92% | 92% | 93% |

TABLE 8
F1 scores for classification experiments with 2 features

| | EROS → MACHO | | EROS → HiTS | |
|---|---|---|---|---|
| Classifier | Original | Transformed | Original | Transformed |
| RF | 51% | 68% | 6% | 27% |
| SVM | 48% | 65% | 0% | 0% |

| | MACHO → EROS | | MACHO → HiTS | |
|---|---|---|---|---|
| Classifier | Original | Transformed | Original | Transformed |
| RF | 47% | 70% | 13% | 50% |
| SVM | 59% | 72% | 1% | 0% |

| | HiTS → EROS | | HiTS → MACHO | |
|---|---|---|---|---|
| Classifier | Original | Transformed | Original | Transformed |
| RF | 1% | 10% | 2% | 7% |
| SVM | 1% | 4% | 1% | 6% |

**Notes.** Column "Original" displays the score obtained when training in the untransformed source dataset and testing on the target dataset. Column "Transformed" shows the score when training on the transformed source dataset and testing on the target dataset.

TABLE 9
F1 scores for classification experiments with 3 features

| | EROS → MACHO | | EROS → HiTS | |
|---|---|---|---|---|
| Classifier | Original | Transformed | Original | Transformed |
| RF | 63% | 73% | 2% | 35% |
| SVM | 75% | 81% | 0% | 1% |

| | MACHO → EROS | | MACHO → HiTS | |
|---|---|---|---|---|
| Classifier | Original | Transformed | Original | Transformed |
| RF | 50% | 62% | 19% | 63% |
| SVM | 68% | 74% | 56% | 60% |

| | HiTS → EROS | | HiTS → MACHO | |
|---|---|---|---|---|
| Classifier | Original | Transformed | Original | Transformed |
| RF | 1% | 26% | 2% | 7% |
| SVM | 0% | 5% | 1% | 4% |

**Notes.** Column "Original" displays the score obtained when training in the untransformed source dataset and testing on the target dataset. Column "Transformed" shows the score when training on the transformed source dataset and testing on the target dataset.

between surveys. Our probabilistic model does not assume a particular classifier and can be used to work with the data of one survey as if it belonged to the other, allowing for the reuse of existing training sets in related domains where no, or not enough, labeled data is available. This will become increasingly important for practical applications as the volume of unlabeled data keeps growing surpassing the rate at which labeled data becomes avail-able. No explicit assumptions are made of the domain shift, which we consider to follow a generalized target shift. We apply our method to simulated data and to three astronomical surveys. First, we do inference in our model to find the transformation parameters. Then, we apply the transformation to the source domain's training set, train a classifier and test in the target domain. Our results show that a significant performance gain in classification can be obtained by adapting a training set with our model, only making use of unlabeled data in both domains.

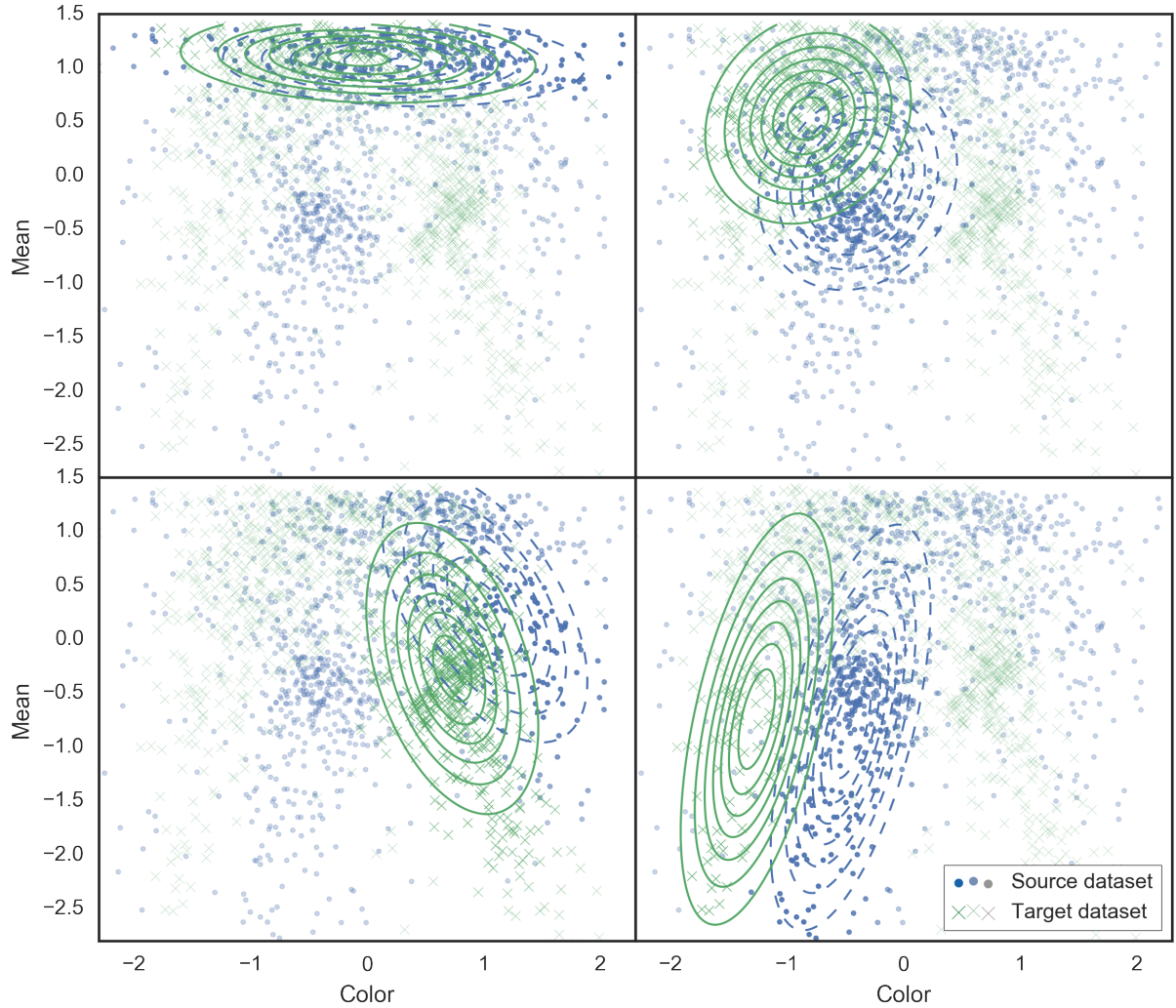

FIG. 9.— Main transformation components from EROS to MACHO. The four components with the highest weights for a 2D transformation from the EROS to the MACHO dataset are shown. The source components are shown with a blue dashed line and the target components with a continuous green line. Point transparency represents the responsibility of the plotted component for that point, with higher opacity representing higher responsibility.

ACKNOWLEDGEMENTS

We acknowledge the support from CONICYT-Chile, through the FONDECYT project number 11140643.

Dataset, Bayesian Model and Parameter Estimation icons designed by Madebyoliver from Flaticon.

TABLE 10
F1 SCORES FOR CLASSIFICATION EXPERIMENTS WITH 4 FEATURES

| Classifier | EROS → MACHO | | EROS → HiTS | |
|---|---|---|---|---|
| | Original | Transformed | Original | Transformed |
| RF | 71% | 84% | 8% | 21% |
| SVM | 84% | 90% | 1% | 1% |

| Classifier | MACHO → EROS | | MACHO → HiTS | |
|---|---|---|---|---|
| | Original | Transformed | Original | Transformed |
| RF | 60% | 70% | 11% | 30% |
| SVM | 78% | 78% | 45% | 44% |

| Classifier | HiTS → EROS | | HiTS → MACHO | |
|---|---|---|---|---|
| | Original | Transformed | Original | Transformed |
| RF | 3% | 11% | 3% | 16% |
| SVM | 0% | 0% | 1% | 1% |

**Notes.** Column "Original" displays the score obtained when training in the untransformed source dataset and testing on the target dataset. Column "Transformed" shows the score when training on the transformed source dataset and testing on the target dataset.

TABLE 11
F1 SCORES FOR CLASSIFICATION USING 2 FEATURES TO TRANSFER FROM EROS TO MACHO

|   | Class | Unadapted RF | Adapted RF | Unadapted SVM | Adapted SVM |
|---|---|---|---|---|---|
| 1 | CEP | 3% | 4% | 7% | 4% |
| 2 | EB | 31% | 37% | 31% | 36% |
| 3 | QSO | 0% | 0% | 0% | 0% |
| 4 | RRLYR | 48% | 79% | 43% | 77% |
| 5 | LPV | 93% | 90% | 94% | 81% |
|   | Weighted Average | 51% | 68% | 48% | 65% |

TABLE 12
F1 SCORES FOR CLASSIFICATION USING 3 FEATURES TO TRANSFER FROM EROS TO MACHO

|   | Class | Unadapted RF | Adapted RF | Unadapted SVM | Adapted SVM |
|---|---|---|---|---|---|
| 1 | CEP | 13% | 0% | 7% | 5% |
| 2 | EB | 36% | 45% | 48% | 59% |
| 3 | QSO | 48% | 40% | 53% | 38% |
| 4 | RRLYR | 63% | 79% | 80% | 90% |
| 5 | LPV | 96% | 95% | 95% | 94% |
|   | Weighted Average | 63% | 73% | 75% | 81% |

TABLE 13
F1 SCORES FOR CLASSIFICATION USING 4 FEATURES TO TRANSFER FROM EROS TO MACHO

|   | Class | Unadapted RF | Adapted RF | Unadapted SVM | Adapted SVM |
|---|---|---|---|---|---|
| 1 | CEP | 7% | 33% | 12% | 44% |
| 2 | EB | 54% | 73% | 71% | 81% |
| 3 | QSO | 19% | 37% | 50% | 60% |
| 4 | RRLYR | 74% | 91% | 89% | 95% |
| 5 | LPV | 95% | 93% | 96% | 94% |
|   | Weighted Average | 71% | 84% | 84% | 90% |

TABLE 14
F1 SCORES FOR CLASSIFICATION USING 2 FEATURES TO TRANSFER FROM EROS TO HiTS

|   | Class | Unadapted RF | Adapted RF | Unadapted SVM | Adapted SVM |
|---|---|---|---|---|---|
| 1 | CEP | 15% | 0% | 15% | 0% |
| 2 | EB | 2% | 1% | 2% | 2% |
| 3 | QSO | 6% | 29% | 0% | 0% |
| 4 | RRLYR | 3% | 6% | 0% | 8% |
|   | Weighted Average | 6% | 27% | 0% | 0% |

TABLE 15
F1 SCORES FOR CLASSIFICATION USING 3 FEATURES TO TRANSFER FROM EROS TO HiTS

|   | Class | Unadapted RF | Adapted RF | Unadapted SVM | Adapted SVM |
|---|---|---|---|---|---|
| 1 | CEP | 0% | 0% | 0% | 0% |
| 2 | EB | 1% | 2% | 1% | 1% |
| 3 | QSO | 2% | 37% | 0% | 0% |
| 4 | RRLYR | 11% | 15% | 5% | 21% |
|   | Weighted Average | 2% | 35% | 0% | 1% |

TABLE 16
F1 SCORES FOR CLASSIFICATION USING 4 FEATURES TO TRANSFER FROM EROS TO HiTS

|   | Class | Unadapted RF | Adapted RF | Unadapted SVM | Adapted SVM |
|---|---|---|---|---|---|
| 1 | CEP | 0% | 0% | 0% | 0% |
| 2 | EB | 1% | 2% | 1% | 1% |
| 3 | QSO | 8% | 21% | 0% | 0% |
| 4 | RRLYR | 10% | 14% | 17% | 16% |
|   | Weighted Average | 8% | 21% | 1% | 1% |

TABLE 17
F1 SCORES FOR CLASSIFICATION USING 2 FEATURES TO TRANSFER FROM MACHO TO EROS

|   | Class | Unadapted RF | Adapted RF | Unadapted SVM | Adapted SVM |
|---|---|---|---|---|---|
| 1 | CEP | 10% | 50% | 5% | 46% |
| 2 | EB | 57% | 68% | 67% | 67% |
| 3 | QSO | 2% | 3% | 0% | 1% |
| 4 | RRLYR | 42% | 73% | 55% | 77% |
| 5 | LPV | 31% | 85% | 43% | 91% |
|   | Weighted Average | 47% | 70% | 59% | 72% |

TABLE 18
F1 SCORES FOR CLASSIFICATION USING 3 FEATURES TO TRANSFER FROM MACHO TO EROS

|   | Class | Unadapted RF | Adapted RF | Unadapted SVM | Adapted SVM |
|---|---|---|---|---|---|
| 1 | CEP | 4% | 3% | 11% | 0% |
| 2 | EB | 59% | 62% | 72% | 72% |
| 3 | QSO | 12% | 16% | 25% | 21% |
| 4 | RRLYR | 45% | 64% | 70% | 80% |
| 5 | LPV | 36% | 73% | 52% | 82% |
|   | Weighted Average | 50% | 62% | 68% | 74% |

TABLE 19
F1 SCORES FOR CLASSIFICATION USING 4 FEATURES TO TRANSFER FROM MACHO TO EROS

|   | Class | Unadapted RF | Adapted RF | Unadapted SVM | Adapted SVM |
|---|---|---|---|---|---|
| 1 | CEP | 14% | 25% | 28% | 7% |
| 2 | EB | 63% | 76% | 80% | 81% |
| 3 | QSO | 16% | 2% | 27% | 5% |
| 4 | RRLYR | 63% | 69% | 82% | 81% |
| 5 | LPV | 32% | 58% | 48% | 76% |
|   | Weighted Average | 60% | 70% | 78% | 78% |

TABLE 20
F1 SCORES FOR CLASSIFICATION USING 2 FEATURES TO TRANSFER FROM MACHO TO HITS

|   | Class | Unadapted RF | Adapted RF | Unadapted SVM | Adapted SVM |
|---|---|---|---|---|---|
| 1 | CEP | 12% | 9% | 18% | 0% |
| 2 | EB | 1% | 2% | 0% | 1% |
| 3 | QSO | 14% | 53% | 1% | 0% |
| 4 | RRLYR | 1% | 2% | 1% | 1% |
|   | Weighted Average | 13% | 50% | 1% | 0% |

TABLE 21
F1 SCORES FOR CLASSIFICATION USING 3 FEATURES TO TRANSFER FROM MACHO TO HITS

|   | Class | Unadapted RF | Adapted RF | Unadapted SVM | Adapted SVM |
|---|---|---|---|---|---|
| 1 | CEP | 17% | 0% | 5% | 0% |
| 2 | EB | 1% | 1% | 1% | 3% |
| 3 | QSO | 20% | 67% | 59% | 63% |
| 4 | RRLYR | 2% | 6% | 9% | 9% |
|   | Weighted Average | 19% | 63% | 56% | 60% |

TABLE 22
F1 SCORES FOR CLASSIFICATION USING 4 FEATURES TO TRANSFER FROM MACHO TO HITS

|   | Class | Unadapted RF | Adapted RF | Unadapted SVM | Adapted SVM |
|---|---|---|---|---|---|
| 1 | CEP | 22% | 3% | 15% | 0% |
| 2 | EB | 1% | 3% | 2% | 2% |
| 3 | QSO | 11% | 38% | 47% | 47% |
| 4 | RRLYR | 16% | 9% | 13% | 10% |
|   | Weighted Average | 11% | 36% | 45% | 44% |

TABLE 23
F1 SCORES FOR CLASSIFICATION USING 2 FEATURES TO TRANSFER FROM HiTS TO EROS

|   | Class | Unadapted RF | Adapted RF | Unadapted SVM | Adapted SVM |
|---|---|---|---|---|---|
| 1 | CEP | 22% | 7% | 38% | 2% |
| 2 | EB | 0% | 13% | 0% | 0% |
| 3 | QSO | 2% | 2% | 2% | 2% |
| 4 | RRLYR | 0% | 7% | 0% | 7% |
|   | Weighted Average | 1% | 10% | 1% | 4% |

TABLE 24
F1 SCORES FOR CLASSIFICATION USING 3 FEATURES TO TRANSFER FROM HiTS TO EROS

|   | Class | Unadapted RF | Adapted RF | Unadapted SVM | Adapted SVM |
|---|---|---|---|---|---|
| 1 | CEP | 15% | 10% | 9% | 2% |
| 2 | EB | 0% | 37% | 0% | 0% |
| 3 | QSO | 2% | 2% | 2% | 2% |
| 4 | RRLYR | 0% | 15% | 0% | 11% |
|   | Weighted Average | 1% | 26% | 0% | 5% |

TABLE 25
F1 SCORES FOR CLASSIFICATION USING 4 FEATURES TO TRANSFER FROM HiTS TO EROS

|   | Class | Unadapted RF | Adapted RF | Unadapted SVM | Adapted SVM |
|---|---|---|---|---|---|
| 1 | CEP | 26% | 19% | 5% | 0% |
| 2 | EB | 4% | 20% | 0% | 0% |
| 3 | QSO | 2% | 2% | 2% | 2% |
| 4 | RRLYR | 0% | 1% | 0% | 0% |
|   | Weighted Average | 3% | 11% | 0% | 0% |

TABLE 26
F1 SCORES FOR CLASSIFICATION USING 2 FEATURES TO TRANSFER FROM HiTS TO MACHO

|   | Class | Unadapted RF | Adapted RF | Unadapted SVM | Adapted SVM |
|---|---|---|---|---|---|
| 1 | CEP | 11% | 6% | 18% | 0% |
| 2 | EB | 6% | 9% | 0% | 0% |
| 3 | QSO | 12% | 11% | 12% | 13% |
| 4 | RRLYR | 0% | 6% | 0% | 8% |
|   | Weighted Average | 2% | 7% | 1% | 6% |

TABLE 27
F1 SCORES FOR CLASSIFICATION USING 3 FEATURES TO TRANSFER FROM HiTS TO MACHO

|   | Class | Unadapted RF | Adapted RF | Unadapted SVM | Adapted SVM |
|---|---|---|---|---|---|
| 1 | CEP | 10% | 0% | 9% | 0% |
| 2 | EB | 0% | 7% | 0% | 0% |
| 3 | QSO | 12% | 14% | 12% | 13% |
| 4 | RRLYR | 2% | 7% | 0% | 5% |
|   | Weighted Average | 2% | 7% | 1% | 4% |

TABLE 28
F1 SCORES FOR CLASSIFICATION USING 4 FEATURES TO TRANSFER FROM HiTS TO MACHO

|   | Class | Unadapted RF | Adapted RF | Unadapted SVM | Adapted SVM |
|---|---|---|---|---|---|
| 1 | CEP | 18% | 9% | 32% | 0% |
| 2 | EB | 10% | 68% | 0% | 0% |
| 3 | QSO | 12% | 8% | 12% | 11% |
| 4 | RRLYR | 0% | 0% | 0% | 0% |
|   | Weighted Average | 3% | 16% | 1% | 1% |